\documentclass[pra,twocolumn,superscriptaddress,preprintnumbers]{revtex4-1}

\usepackage{amsthm}
\usepackage{amsmath}
\usepackage{mathrsfs}
\usepackage{amssymb}
\usepackage{wasysym}
\usepackage{graphicx}
\usepackage{varwidth}
\usepackage{url}
\usepackage{dsfont}
\usepackage{float}
\usepackage{bm}
\usepackage{ifthen}
\usepackage[usenames,dvipsnames]{color}
\usepackage{mathrsfs}
\usepackage{hyperref}
\usepackage{float}
\usepackage{afterpage}
\usepackage[caption=false]{subfig}
\usepackage{textcomp}

\newsavebox{\twosubbox}
\usepackage{physics}
\usepackage{xcolor}
\usepackage{fancyhdr}
\usepackage{ulem}

\usepackage{enumitem}

\usepackage{algorithm,algpseudocode}
\algnewcommand{\LeftComment}[1]{\Statex \hspace{3em}\(\triangleright\) #1}

\usepackage{amsthm}

\theoremstyle{definition}

\begin{document}
\preprint{CaltechAUTHORS:20230316-190343094; FERMILAB-PUB-23-120-ETD}
\title{Comment on ``Comment on `Traversable wormhole dynamics\\
on a quantum processor'"}
\begin{abstract}
We observe that the comment of 
 \citet{kobrin2023comment}
is consistent with \citet{jafferis2022traversable} on key points: i) the microscopic mechanism of the experimentally observed teleportation is size winding
and ii) the system thermalizes and scrambles at the time of teleportation. These properties are consistent with a gravitational interpretation of the teleportation dynamics, as opposed to the late-time dynamics. The objections of \citet{kobrin2023comment} concern counterfactual scenarios outside of the  experimentally implemented protocol.

\begin{enumerate}[wide, leftmargin = 0pc, rightmargin = 5pc]
    \item The first scenario of \cite{kobrin2023comment} asks about times after the teleportation: they extend the dynamics via single-sided evolution to conclude that the learned Hamiltonian does not thermalize. We find that wormhole teleportation persists despite the addition of a non-commuting perturbation that reduces single-sided revivals.
    Moreover, we show thermalization under the learned eternal traversable wormhole Hamiltonian, which, like the experimentally implemented protocol, couples the left and right systems.
    \item The second scenario of \cite{kobrin2023comment} asks about teleporting different fermions: they claim no size winding occurs on untrained fermions. We find all fermions are teleported by size winding albeit at different times. A strongly gravitational signature emerges: fermions that thermalize more slowly exhibit size winding at later times, consistent with a holographic interpretation of less massive fermions  having a more delocalized wavepacket, taking them  longer to traverse the wormhole.
    \item The third scenario of \cite{kobrin2023comment} asks about different Hamiltonians: they find similar behavior in random commuting Hamiltonians. Besides identifying technical issues 
    in their analysis, we note that an ensemble of similar Hamiltonians to the learned Hamiltonian should exist and exhibit similar properties. We also show the commuting structure is unrelated to the presence of gravitational physics by adding a large non-commuting term and finding that size winding is preserved.
\end{enumerate}
\end{abstract}

\author{Daniel Jafferis}
\thanks{These two authors contributed equally.}
\affiliation{Center for the Fundamental Laws of Nature, Harvard University, Cambridge, MA, USA}

\author{Alexander Zlokapa}
\thanks{These two authors contributed equally.}
\affiliation{Center for Theoretical Physics, Massachusetts Institute of Technology, Cambridge, MA, USA}
\affiliation{Division of Physics, Mathematics and Astronomy, Caltech, Pasadena, CA, USA}
\affiliation{Alliance for Quantum Technologies (AQT), California Institute of Technology, Pasadena, CA, USA}
\affiliation{Google Quantum AI, Venice, CA, USA}

\author{Joseph D. Lykken}
\affiliation{Fermilab Quantum Institute, Fermi National Accelerator Laboratory, Batavia, IL, USA}

\author{David K. Kolchmeyer}
\affiliation{Center for the Fundamental Laws of Nature, Harvard University, Cambridge, MA, USA}

\author{Samantha I. Davis}
\affiliation{Division of Physics, Mathematics and Astronomy, Caltech, Pasadena, CA, USA}
\affiliation{Alliance for Quantum Technologies (AQT), California Institute of Technology, Pasadena, CA, USA}

\author{Nikolai Lauk}
\affiliation{Division of Physics, Mathematics and Astronomy, Caltech, Pasadena, CA, USA}
\affiliation{Alliance for Quantum Technologies (AQT), California Institute of Technology, Pasadena, CA, USA}

\author{Hartmut Neven}
\affiliation{Google Quantum AI, Venice, CA, USA}

\author{Maria Spiropulu}
\affiliation{Division of Physics, Mathematics and Astronomy, Caltech, Pasadena, CA, USA}
\affiliation{Alliance for Quantum Technologies (AQT), California Institute of Technology, Pasadena, CA, USA}

\maketitle

\section{Introduction}
The purpose of \citet{jafferis2022traversable} was to perform quantum many-body teleportation with a traversable wormhole interpretation. This requires satisfying key properties \emph{at the time of the teleportation}. Before the interaction between the left and right systems is applied, the transmitted fermions must thermalize and scramble in the left system, as defined by the decay of two-point and out-of-time-order correlators. The fermions must spread through the left system via operator growth with the particular phase coherence described by size winding. Applying the interaction must then reverse the direction of the size winding, ensuring that time evolution causes the fermions to unthermalize and unscramble on the right system. We observe that the numerical results of \citet{kobrin2023comment} are fully consistent with this picture.

\citet{kobrin2023comment} raises objections over a counterfactual scenario regarding the dynamics of a single-sided system with no interaction to perform the traversable wormhole protocol. We show that introducing a large non-commuting perturbation that damps revivals of the single-sided system at late times has minimal impact on transmission dynamics at the time of teleportation; this shows a decoupling between the issue of late-time dynamics and a gravitational interpretation of the teleportation. Moreover, we consider a gravitationally interpretable counterfactual scenario: the evolution of the coupled left-right system, corresponding to an eternal traversable wormhole \cite{maldacena2018eternal}. The wormhole teleportation protocol proposed in \cite{gao2021traversable} and implemented in our experiment is strictly equivalent to a single Trotter step of time evolution under the eternal traversable wormhole Hamiltonian. Due to the left-right interaction, we find that this coupled system thermalizes at high temperature as expected from the gravitational interpretation \cite{maldacena2018eternal}.

A second counterfactual discussed by \cite{kobrin2023comment} is the teleportation of different fermions: they claim that the learned Hamiltonian is biased to have good size winding only on the two fermions implemented in the experiment. This objection is not directly relevant to general holographic models, which can have very different properties across different fermions; not all fermions must have size winding. Nevertheless, in the learned Hamiltonian, we find that all fermions show size winding at times $2 \lesssim t \lesssim 5$; their claim is an artifact of only analyzing size winding at $t=2.8$. We find that fermions that thermalize more slowly in the eternal traversable wormhole Hamiltonian achieve good size winding at later times. This is a gravitationally meaningful feature associated with different masses across fermions: a lighter fermions has a delocalized wave packet that causes it to thermalize more slowly and take longer to traverse the wormhole, hence exhibiting size winding later.   A late time corresponds to a boost in the near-horizon region, which makes the wavepacket more localized via length contraction.  We view this time dependence as evidence for a holographic interpretation of the learned Hamiltonian. We note that the size winding behavior ($2 \lesssim t \lesssim 5$) occurs before the model experiences revivals from its commuting structure. To ensure gravitational dynamics at later times ($t \gtrsim 10$), one needs to apply a perturbation such as a left-right coupling or non-commuting Floquet evolution, as described in the previous paragraph.

Finally, \citet{kobrin2023comment} comment on the properties of random Hamiltonians of the same structure as the learned Hamiltonian \cite{jafferis2022traversable}. Below, we note technical issues with this analysis and find that a strict minority of random Hamiltonians have as good size winding as the learned Hamiltonian at the time of teleportation. Crucially, however, we see no contradiction in non-bulk explanations of perfect size winding. We expect similar Hamiltonians to exhibit similar properties. Moreover, the commuting structure of the Hamiltonian is not central to achieving size winding properties. We provide a large perturbation of the learned Hamiltonian that thermalizes but has similar size winding and teleportation behavior, suggesting that the commuting structure is irrelevant to the presence of gravitational physics.

\section{Gravitational dynamics after teleportation}

In the setting of a traversable wormhole, the teleportation time of the coupled system defines the relevant timescale of the dynamics. On this timescale, the two-point and out-of-time-order correlators decay for individual fermions and for the average over all fermions. Our usage of ``thermalization" and ``scrambling" in the main text of \cite{jafferis2022traversable} reflects the properties of the learned Hamiltonian on the timescale of the teleportation ($t\approx 2.8$). At $t=2.8$, we found that the two Majorana fermions being teleported each spread onto 8 operators, with the largest operator of size 5. (In the absence of the commuting structure, each fermion would spread onto  36 operators in an $N=7$ SYK model.)

To build a minimal example of a system that is interpretable as a traversable wormhole, the learning procedure coarse-grained out details concerning behavior after the teleportation time. Consequently, different counterfactual scenarios that occur after the teleportation can be subject to different interpretations. \citet{kobrin2023comment} considered a direct extension of system dynamics: evolving the single-sided Hamiltonians independently. Due to the commuting structure of the Hamiltonian, the two-point function exhibits revivals that prevent thermalization and hence prevents interpreting each single-sided system as a black hole. Since these revivals occur by time $t\approx 10$, dynamics such as the mutual information can show non-gravitational behavior at times $t_0 + t_1 \gtrsim 10$.

The revivals of the single-sided system after the time of teleportation can be straightforwardly suppressed by introducing a periodic non-commuting perturbation, producing behavior more similar to the SYK model as we shall demonstrate in Sec.~\ref{sec:comm}. We will find that although the late-time dynamics are now governed by a non-commuting Hamiltonian, the behavior of the system at the time of teleportation ($t\approx 2.8$) is unchanged. In particular, the teleportation dynamics at $t\approx 2.8$ remain consistent with the expected gravitational signature. This shows that the late-time thermalization of the model --- satisfying the single-sided counterfactual proposed by~\cite{kobrin2023comment} --- is a feature that is decoupled from the existence of gravitational dynamics at early times. For a full discussion of this decoupling, see Sec.~\ref{sec:comm}.

\begin{figure}[H]
  \centering
  \includegraphics[width=\columnwidth]{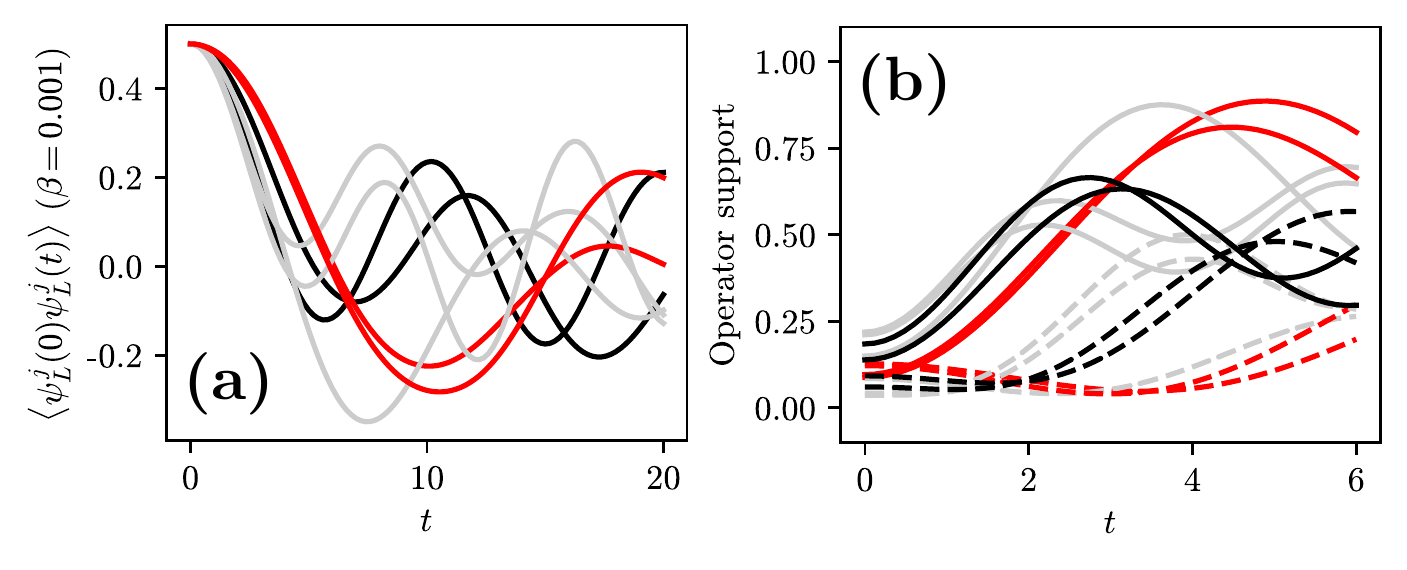}
  \caption{\textbf{Thermalization and operator growth.} \textbf{(a)}, The left-left two-point function of the coupled eternal traversable wormhole Hamiltonian ($\beta=0.001$). The two fermions teleported in the traversable wormhole experiment are shown in black ($\psi^1, \psi^2$); the two fermions identified by \cite{kobrin2023comment} to have poor size winding are shown in red ($\psi^4, \psi^7$). Fermions $\psi^4,\psi^7$ are the slowest to thermalize. \textbf{(b)}, The single-sided operator growth of individual fermions. Solid lines show support $p(l)$ (Eq.~\ref{eq:p}) over operators of size $l=3$, and dashed lines show support over operators of size $l=5$. Fermions $\psi^4$ and $\psi^7$ grow the slowest, indicating that they traverse the wormhole later.}
  \label{fig:f1}
\end{figure}

\begin{figure*}[t]
  \centering
  \includegraphics[width=\textwidth]{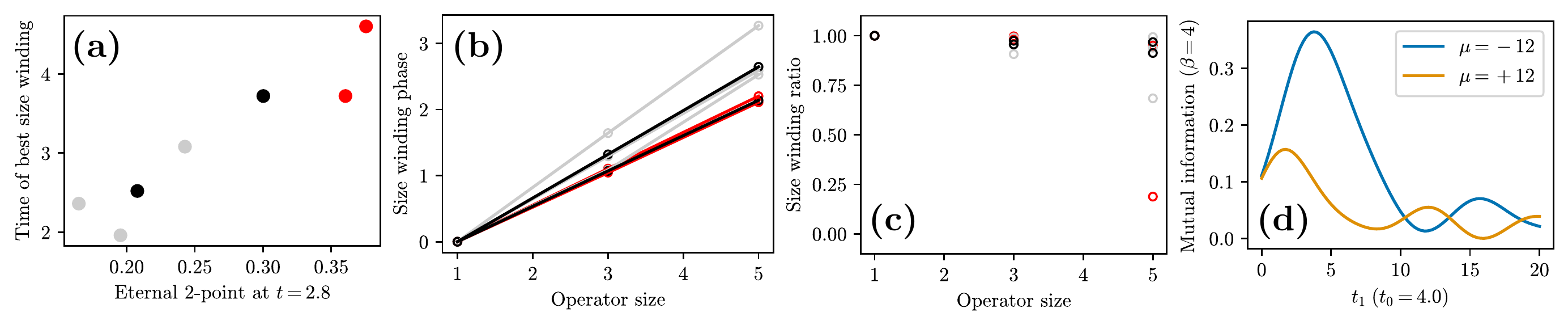}
  \caption{\textbf{Size winding of individual fermions.} \textbf{(a)}, The time of best size winding vs. the value of the two-point function shown in Fig.~\ref{fig:f1}(a) at $t=2.8$. Fermions that thermalize more slowly have good size winding at later times. As in Fig.~\ref{fig:f1}, fermions $\psi^4$ and $\psi^7$ are shown in red, and fermions $\psi^1$ and $\psi^2$ are shown in black. \textbf{(b)}, The size winding phase $\arg q(l)$ (Eq.~\ref{eq:q}) for each of the fermions at the time of their best size winding. All fermions achieve near-linear phase dependence. \textbf{(c)}, The ratio $|q(l)|/p(l)$ for all fermions. \textbf{(d)}, The wormhole teleportation protocol using fermions $\psi^4$ and $\psi^7$ and applying the interaction at $t=4$. The fermions with the worst size winding (previously colored red) show a clear mutual information peak asymmetry at the time of teleportation, showing that they teleport by size winding.}
  \label{fig:f2}
\end{figure*}

Here, we examine a different counterfactual scenario that leads to meaningful gravitational behavior. We observe that the entire coupled system
\begin{align*}
    H_\mathrm{tot} &= H_L + H_R + \mu_\mathrm{MQ} H_\mathrm{int}, \; H_\mathrm{int} = i\sum_j \psi_L^j \psi_R^j
\end{align*}
behaves as expected from the eternal traversable wormhole interpretation \cite{maldacena2018eternal}. Note that the protocol implemented in the traversable wormhole experiment is equivalent, at the circuit level, to a first-order Trotterization of $e^{-iH_\mathrm{tot}t}$ with a single Trotter step. The Hamiltonian $H_\mathrm{tot}$ exhibits operator growth and thermalizes at high temperature (Fig. \ref{fig:f1}a), as is expected of the eternal traversable wormhole. No revivals occur in the two-point function, even at late times, due to the presence of $H_\mathrm{int}$. At low temperature, since the ground state of $H_\mathrm{tot}$ is the thermofield double state and has an $O(1)$ gap to the first excited state, the two-point function appropriately exhibits larger oscillations.

In Fig.~\ref{fig:f1}a, a diversity of decay rates of the two-point function is seen across different fermions; highlighted in red are the two fermions that decay the slowest. The gravitational interpretation of thermalization rates is that each fermion corresponds to a different mass. Since the wave packet of a lighter fermion is more spread out, its two-point function decays more slowly.

We check this behavior by examining the single-sided system $H_L$. Inserting a fermion onto the TFD and time evolving under $H_L$ should result in operator growth, with lighter fermions growing more slowly. In Fig.~\ref{fig:f1}b, we see that the fermions that exhibit the slowest operator growth in $H_L$ are precisely the same fermions that thermalize the slowest in $H_\mathrm{tot}$. When examining size winding in the following section, we shall define the precise measure of operator growth (Eq.~\ref{eq:p}) and observe further behavior consistent with the gravitational interpretation of fermions with different masses.

\section{Size winding}

The size winding analysis included in the main text of \cite{jafferis2022traversable} depicts the first fermion of the learned Hamiltonian. Here, we also show the size winding of all fermions at different times.

A holographic dual need not have size winding across all fermions; if a subset of fermions have additional non-gravitational interactions, they would not necessarily exhibit size winding. Consequently, the absence of size winding on fermions other than those teleported have no known implication for the holographic dual of the teleported fermions.

\begin{figure*}[t]
  \centering
  \includegraphics[width=0.78\textwidth]{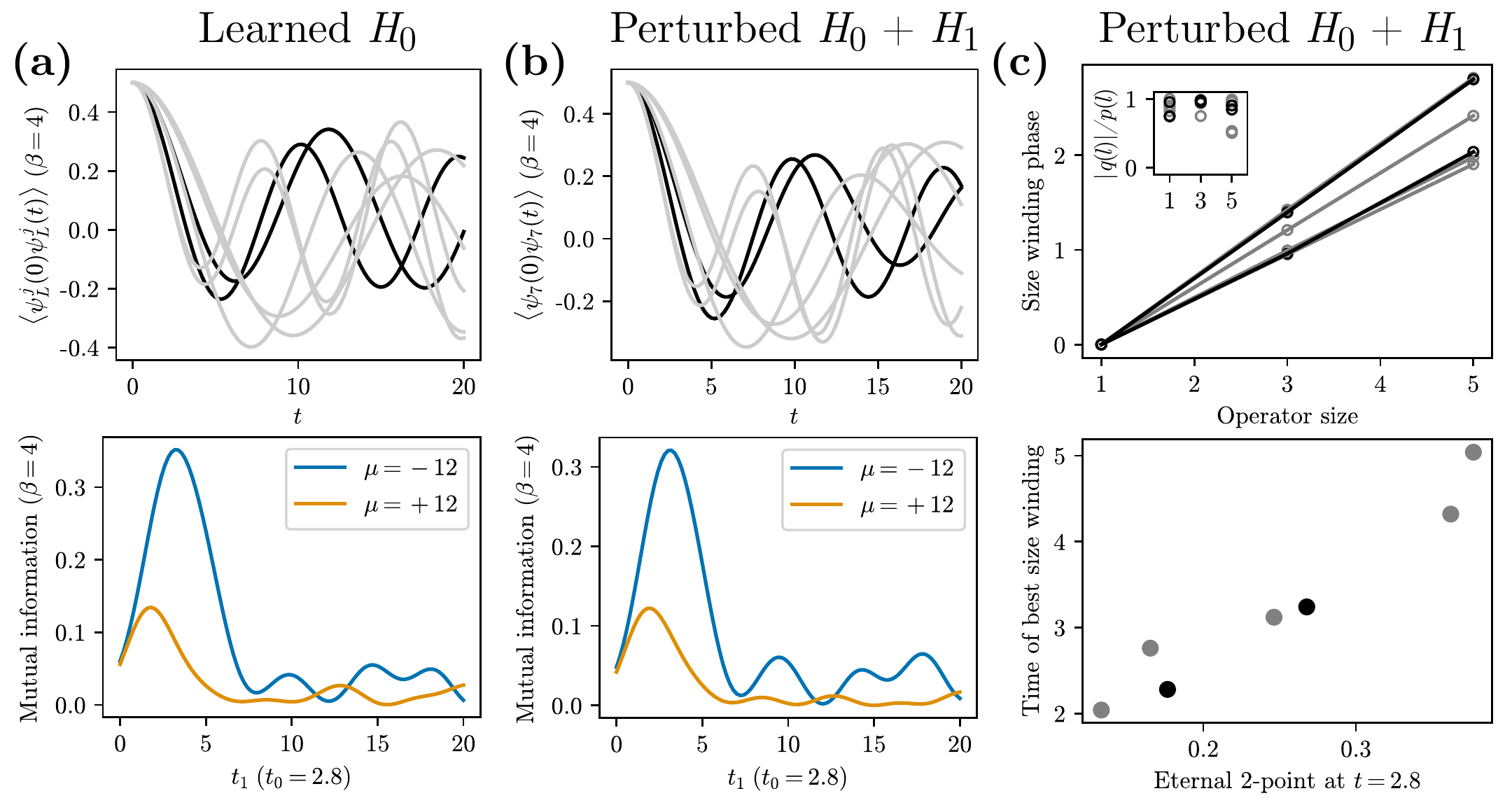}
  \caption{\textbf{Behavior of perturbed Hamiltonian.} \textbf{(a)}, Two-point function (top) and mutual information dynamics (bottom) of individual fermions in the learned Hamiltonian ($\psi^1,\psi^2$ in black; other fermions in gray). \textbf{(b)}, Two-point function (top) and mutual information dynamics (bottom) of individual fermions in the perturbed Hamiltonian (same colors). Compared to (a), the two-point function of $H_0+H_1$ does not experience as large revivals due to a non-commuting term and shows slightly damped oscillatory behavior. The mutual information dynamics preserve the gravitational signature of asymmetry in the interaction sign. \textbf{(c)}, The perturbed Hamiltonian's size winding phases $\arg q(l)$ (top) and ratios $|q(l)|/p(l)$ (inset) shown at the time of best size winding (bottom). All fermions show size winding at different times, which is seen in (b) to be sufficient to generate a gravitational teleportation signature. Lighter fermions thermalize more slowly and traverse the wormhole at later times.}
  \label{fig:f3}
\end{figure*}

Attaching different masses to the fermions has an implication for size winding. Since lighter fermions traverse the wormhole more slowly, the event of traversing the wormhole occurs later. Consequently, the microscopic mechanism of size winding should occur at later times. In Fig.~\ref{fig:f2}a, we see this expectation holds for the learned Hamiltonian.

Size winding decomposes a time-evolved fermion $\psi_L^j(t)$ over Majorana strings of size $|P|$
\begin{align*}
    \rho_\beta^{1/2}\psi_L^j(t) = \sum_P c_P(t) \psi_L^P,
\end{align*}
where the support of the fermion on operators of size $l$ is given by
\begin{align}
\label{eq:p}
    p(l, t) = \sum_{|P|=l} |c_P(t)|^2
\end{align}
and the winding size distribution is given by
\begin{align}
\label{eq:q}
    q(l, t) = \sum_{|P|=l} c_P(t)^2.
\end{align}
We measure the quality of size winding by evaluating the linearity of size winding phases $\arg q(l, t)$; in perfect size winding at time $t$, $q(l, t)$ is linear in $l$. In Fig.~\ref{fig:f2}b, we show the phases of each of the fermions at the time they are most linear. This demonstrates that all fermions eventually achieve near-perfect linearity \footnote{We measure the quality of size winding via the standard deviation of the difference in adjacent values of $\arg q(l)$. The time of best size winding is obtained by identifying the first local minimum in the standard deviation of size winding phases after $t=0$; we also require the standard deviation to fall below a reasonable threshold (0.8).}.

\citet{kobrin2023comment} observe that fermions $\psi^4$ and $\psi^7$ (colored red) have poor size winding at $t=2.8$. We see in Fig.~\ref{fig:f2}b that those fermions achieve near-perfect linearity of size winding phases at slightly later times ($t\approx 4$, shown in Fig.~\ref{fig:f2}a). Those fermions also thermalize more slowly in the coupled system (Fig.~\ref{fig:f1}a) and experience slower operator growth in the single-sided system (Fig.~\ref{fig:f1}b). This is consistent with interpreting them as taking longer to traverse the wormhole. While the ratio $|q(l)|/p(l)$ is not unity for all fermions (Fig.~\ref{fig:f2}c), it is sufficiently large to achieve a teleportation signal with sign dependence on the coupling for fermions $\psi^4$ and $\psi^7$ (Fig.~\ref{fig:f2}d).

\section{Commuting Hamiltonian structure}
\label{sec:comm}
In \cite{jafferis2022traversable}, we used a learning procedure to identify a minimal example of a Hamiltonian with traversable wormhole dynamics. The learning procedure involves a free parameter that controls the sparsity of the SYK model, generating an ensemble that includes non-commuting Hamiltonians but preserves gravitationally relevant properties.

\emph{A priori}, it was not known that a small commuting Hamiltonian (Eq.~\ref{eq:H}) would exhibit behavior such as size winding. After this finding, it may be expected that additional Hamiltonians of similar form have similar properties. Indeed, while the learning procedure may aid in the discovery of Hamiltonians with gravitational behavior, there always exists a non-bulk explanation of its behavior. For this reason, the presence of size winding in Hamiltonians explicitly constructed to be similar to the learned Hamiltonian is unsurprising; it has no relevance in determining if the learned Hamiltonian behaves gravitationally.

Here, we examine if the commuting structure of the Hamiltonian is intrinsically related to its size winding properties and dynamics. We take the learned Hamiltonian
\begin{align}
\label{eq:H}
    H_0 = &-0.36 \psi^1\psi^2\psi^4\psi^5 + 0.19 \psi^1\psi^3\psi^4\psi^7 \nonumber\\
    &- 0.71\psi^1\psi^3\psi^5\psi^6 + 0.22\psi^2\psi^3\psi^4\psi^6 \nonumber \\
    &+ 0.49\psi^2\psi^3\psi^5\psi^7,
\end{align}
and add a perturbation with coefficient roughly equal to the median coefficient in $H_0$,
\begin{align}
\label{eq:pert}
    H_1 = 0.3\psi^1\psi^2\psi^3\psi^5,
\end{align}
which forces the system to thermalize at long timescales but does not significantly modify the two-point function or mutual information dynamics at the timescale $t=2.8$ of the wormhole teleportation (Fig.~\ref{fig:f3}ab). In particular, the two fermions $\psi^1, \psi^2$ that traverse the wormhole do not exhibit revivals in the perturbed system $H_0+H_1$, unlike the revivals seen in just $H_0$. The perturbed system $H_0+H_1$ shows size winding (Fig.~\ref{fig:f3}c) sufficient to produce an asymmetric teleportation signal.

To emphasize the independence of 1) exhibiting gravitational physics at the time of teleportation and 2) thermalizing at late times, we show that applying the non-commuting perturbation in Trotterized time evolution still produces the mutual information dynamics characteristic of a gravitational signal. We prepare the thermofield double state using only the learned Hamiltonian, but evolve under a time-dependent Floquet Hamiltonian that alternates between the perturbation ($H_1$) and the commuting system ($H_0$), with the alternation occurring in intervals of $\Delta t=2.8$. A fermion explores a space of 12 operators under the Floquet Hamiltonian instead of 8 operators under $H_0$.

\begin{figure}[H]
  \centering
  \includegraphics[width=\columnwidth]{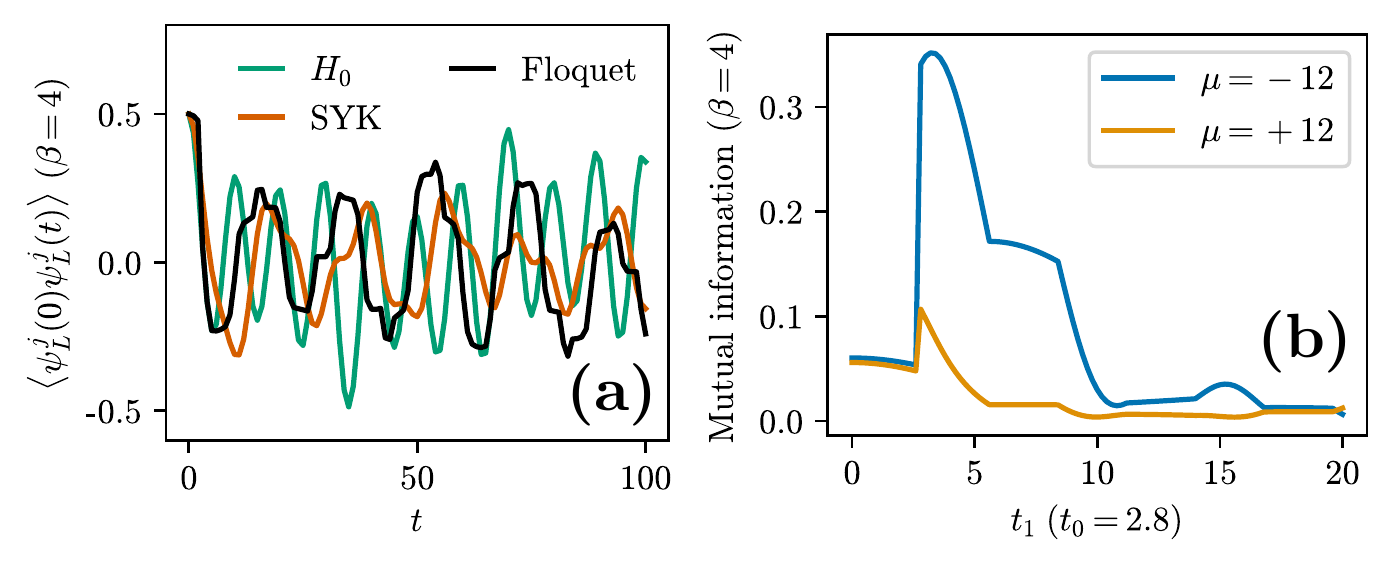}
  \caption{\textbf{Behavior of Floquet Hamiltonian.} \textbf{(a)}, Two-point function of the single-sided system under alternating evolution by $H_0$ and $H_1$ (Floquet), compared to $H_0$ and an SYK model. \textbf{(b)}, Mutual information dynamics of the Floquet Hamiltonian with fixed injection time $-t_0=-2.8$. The system shows gravitational behavior despite the insertion of periodic non-commuting evolution.}
  \label{fig:f4}
\end{figure}

We compare the dynamics of the Floquet Hamiltonian to the learned Hamiltonian $H_0$ and an $N=7$ SYK model (35 terms) in Fig.~\ref{fig:f4}a.  In 
Fig.~\ref{fig:f4}b we show the asymmetry in mutual information in the Floquet case, 
demonstrating decoupling between the gravitational behavior at the time of teleportation from late-time dynamics. 
The perturbation would satisfy the late-time single-sided counterfactual argued by \citet{kobrin2023comment}; however, it has no significant effect on the physics at the time of the teleportation.

This example demonstrates that adding a large non-commuting term does not introduce any meaningful change to the gravitationally interesting regime ($t\approx 2.8$, $\mu=-12$). It suggests that the commuting structure of the learned Hamiltonian is not intrinsic to the relevant behavior we observe at the time of teleportation.

\citet{kobrin2023comment} comment on two methods of generating Hamiltonians similar to Eq.~\ref{eq:H}. First, they consider randomizing the coefficients by sampling from a normal distribution; second, they randomize both the coefficients and the choice of terms, such that they obtain a 7-fermion 5-term commuting Hamiltonian. We point out that these are strictly equivalent procedures: there exists a unique 7-fermion 5-term commuting Hamiltonian, up to relabeling of fermions, given by
\begin{align*}
    H =\;&\alpha_1 \psi^1\psi^2\psi^3\psi^4 + \alpha_2 \psi^1\psi^2\psi^5\psi^6 \\
    &+ \alpha_3 \psi^3\psi^4\psi^5\psi^6 + \alpha_4 \psi^1\psi^3\psi^5\psi^7 \\
    &+ \alpha_5 \psi^2\psi^4\psi^5\psi^7.
\end{align*}
Hence, ``randomizing coefficients and terms" is precisely the same as simply randomizing coefficients.

The authors of \cite{kobrin2023comment} claim that the quality of size winding ``over all operators resembles that of generic random small-size fully commuting models." We reproduce the method of generating random Hamiltonians reported by \cite{kobrin2023comment} and observe that at $t=2.8$, from 1000 random instances, 29\% of random models have as good size winding on the best two fermions and only 3\% have as good size winding on all fermions, as measured by linearity of $\arg q(l)$. Measured by the variable $\chi$ proposed by \cite{kobrin2023comment}, these numbers are 25\% and 2\% respectively. This 
contradicts the claim of \cite{kobrin2023comment}, which analyzed size winding at $t=2.8$.

Setting aside these technical issues with the analysis of \cite{kobrin2023comment}, we reiterate that the presence of similar Hamiltonians with similar properties does not have any bearing on a gravitational interpretation of the learned Hamiltonian $H_0$.

This work is supported by the Department of Energy Office of High Energy Physics QuantISED program grant SC0019219 on Quantum Communication Channels for Fundamental Physics. This manuscript has been authored by Fermi Research Alliance, LLC under Contract No. DE-AC02-07CH11359 with the U.S. Department of Energy, Office of Science, Office of High Energy Physics.

\bibliography{scibib}

\end{document}